\documentclass[twoside]{article}

\usepackage{PRIMEarxiv}
\usepackage{amsmath}
\usepackage{todonotes}
\usepackage[utf8]{inputenc} % allow utf-8 input
\usepackage[T1]{fontenc}    % use 8-bit T1 fonts

\PassOptionsToPackage{colorlinks=true, allcolors=blue}{hyperref}

\usepackage{hyperref}       % hyperlinks
\usepackage{url}            % simple URL typesetting
\usepackage{booktabs}       % professional-quality tables
\usepackage{amsfonts}       % blackboard math symbols
\usepackage{nicefrac}       % compact symbols for 1/2, etc.
\usepackage{microtype}      % microtypography
\usepackage{natbib}         % for \citep and \citet
\usepackage{lipsum}
\usepackage{fancyhdr}       % header
\usepackage{subcaption}
\usepackage{mathtools}
\usepackage[normalem]{ulem}
\usepackage[dvipsnames]{xcolor}
\usepackage{fancyvrb}
\RecustomVerbatimCommand{\VerbatimInput}{VerbatimInput}%
{%fontsize=\scriptsize,
 frame=lines,  % top and bottom rule only
 framesep=2em, % separation between frame and text
 rulecolor=\color{Gray},
}

\usepackage[english]{babel}

\usepackage{graphicx}

\usepackage{listings}

\usepackage{minted}
\setminted{
  fontsize=\small,
  frame=lines,
  framesep=2mm,
  breaklines=true,
  tabsize=2,
}

\usepackage{color,soul}
\usepackage{caption}
\def\CO2{CO\textsubscript{2}}
\def\H2O{H\textsubscript{2}O}

\usepackage{multirow}

\usepackage{placeins}
\usepackage{tikz}
\usetikzlibrary{positioning, arrows.meta, calc, decorations.pathreplacing, bending, fit, backgrounds}

\definecolor{userblue}{HTML}{2563EB}
\definecolor{agentgreen}{HTML}{1F9D55}
\definecolor{toolorange}{HTML}{E08A1E}
\definecolor{foundgray}{HTML}{475569}
\definecolor{knowpurple}{HTML}{7C3AED}
\definecolor{flowcyan}{HTML}{0EA5E9}

\title{Agents4GEOS: Agentic Platform for Open-Source Multi-physics Simulation}

\author{%
  Adriano M. A. Côrtes$^{1}$\thanks{Corresponding author: adricortes@cos.ufrj.br} \and
  Roberto M. Velho$^{1}$ \and
  Fernando A. Rochinha$^{1}$ \and
  Alvaro L. G. A. Coutinho$^{1}$ \and Mauricio Araya-Polo$^{2}$\thanks{Corresponding author: mauricio.araya@totalenergies.com} \and Hervé Gross$^{2}$
}

\date{}

\begin{document}

\maketitle
\vspace{-25pt}
\begin{center}
    \footnotesize
    $^{1}$Federal University of Rio de Janeiro, Brazil.\\
    $^{2}$TotalEnergies EP Research \& Technologies USA, Houston, USA\\

\end{center}

\vspace{15pt}

\begin{abstract}
Multi-physics simulations are essential for understanding and monitoring intricate subsurface processes such as \CO2 storage. Their computational demands call for surrogate models and, for unstructured meshes, Graph Neural Networks (GNNs) are natural candidates. The main bottleneck in developing them is generating and managing the large, physically consistent simulation datasets required for training.

To address this challenge, we present Agents4GEOS, an AI-agent framework built on the Model Context Protocol (MCP) that provides 52 domain-aware tools for natural-language-driven workflows with GEOS, an open-source multi-physics simulator. The agent facilitates input-file creation, mesh inspection, fluid-property computation, and result post-processing.

Through human-curated skills and fresh-context subagents coordinated by an orchestrator, the system executes complex workflows, evaluates simulation outputs, diagnoses issues, and suggests improvements, grounding every quantity in actual computation. By automating routine tasks, Agents4GEOS allows domain experts to focus on the most challenging aspects of their work.
\end{abstract}

%%%%%%%%%%%%%%%%%%%%%%%%%%%%%%%%%
\section{Introduction}

The increasing complexity of computational science workflows, spanning numerical simulation, parameter estimation, and model discovery, has driven renewed interest in autonomous agents capable of coordinating heterogeneous tasks without continuous human intervention. Recent advances in large language models (LLMs) have opened a new paradigm for such orchestration: rather than serving merely as conversational interfaces, LLMs are now being positioned as reasoning engines that can plan, decompose, and delegate subtasks across specialized computational tools. This vision has been explicitly articulated in the context of computer-aided engineering, where LLM-empowered systems can bridge the gap between domain expertise and simulation infrastructure, enabling more adaptive and automated design workflows \cite{guo2026large}.

A particularly active frontier is the coupling of agentic systems with physics-based and data-driven solvers for partial differential equations (PDEs). Recent work has proposed frameworks in which agents navigate the latent spaces of foundation models to explore families of PDE solutions, enabling efficient parameterized simulations without re-running full solvers for every configuration \cite{vishwasrao2026agentic}. This capability is directly relevant to reservoir engineering, where many-query problems, such as uncertainty quantification, history matching, and production optimization, require repeated evaluations of subsurface flow models across large parameter spaces. The computational cost of these workflows has historically been prohibitive, making agent-driven surrogate exploration a compelling alternative.

Related work on multi-agent architectures highlights the potential for emergent problem-solving behavior when specialized agents collaborate. Collaborative multi-agent systems have been shown to accelerate discovery in scientific machine learning by distributing hypothesis generation, model training, and validation across coordinated agents \cite{jiang2026agenticsciml}. Complementarily, hierarchical agentic teams designed specifically to evolve numerical algorithms demonstrate that orchestration hierarchies, where high-level planners delegate to low-level executors, can yield robust and adaptive solvers \cite{toscano2025athena}. In \cite{ramatullayev2026multiagentic} a multi-agentic Generative AI framework integrating LLMs with domain-specific engines is introduced to automate reservoir simulation workflows, targeting model compliance, insight generation, and well placement optimization, demonstrating substantial reductions in manual effort and improved decision-making efficiency in field development planning. Also, in \cite{onishi2026agentic} an agentic AI framework built on NVIDIA's accelerated computing platform is introduced to automate subsurface reservoir simulation workflows, combining a reservoir simulation assistant for daily tasks with a multi-agent squad for complex studies such as history matching and well placement optimization, effectively eliminating operational dead time through continuous 24/7 simulation loops.
Together, these works suggest a convergent direction: agentic orchestration frameworks that combine planning intelligence with domain-specialized numerical tools hold significant promise for transforming reservoir simulation workflows, from initial model construction through to field-scale forecasting and decision support.

Within this landscape, we are developing \emph{Agents4GEOS}, an agentic system that lowers the barrier to building physically consistent GEOS\footnote{\url{https://github.com/GEOS-DEV/GEOS}} simulations and, ultimately, the large datasets that data-driven surrogates require \cite{ju2024learning, luna2026benchmarking, lu2026learninginferringmultiphaseflow}. GEOS \cite{settgast2024geos} is an open-source, performance-portable multiphysics reservoir simulator with native support for unstructured meshes, which makes it an ideal data generator for graph-based surrogates. Its expressiveness, however, comes at a cost: a single simulation is described by an XML input file that typically spans several hundred lines of cross-referenced parameters distributed across solvers, constitutive models, meshes, geometry, boundary conditions, events, and outputs. Authoring and debugging these files by hand is tedious and error-prone, constituting a steep entry barrier for new users and a recurring bottleneck in the many-query workflows on which surrogate construction depends.

\begin{figure*}[ht!]
    \centering
    \includegraphics[width=1.0\linewidth]{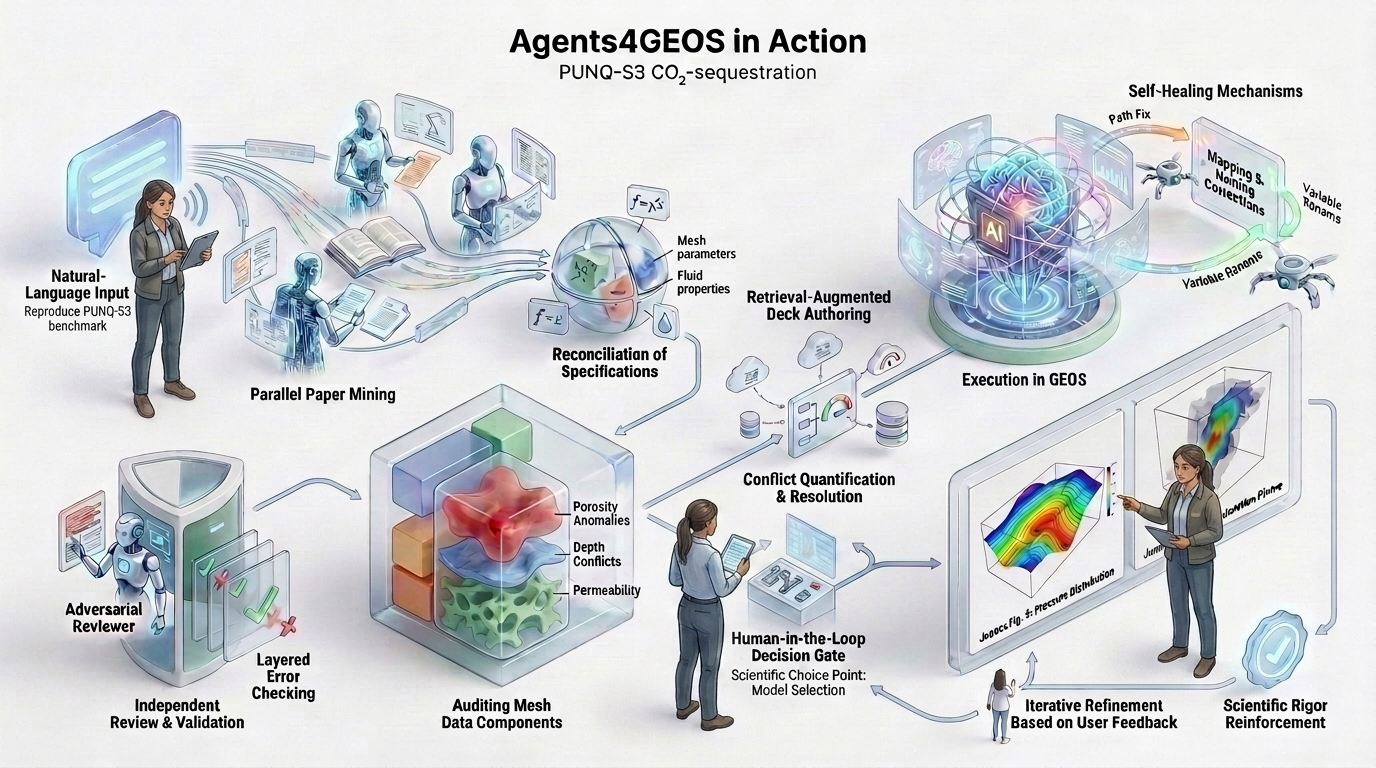}
    \caption{Agents4GEOS reframes the task of simulation data generation as a natural-language dialogue. The user describes the desired simulation in plain English, and a team of specialized agents collaborates to produce a schema-valid, physically sane GEOS XML file, backed by real fluid-property and meshing computations rather than by free-form text generation.}
    \label{fig:agents_scheme}
\end{figure*}

Agents4GEOS reframes this task as a natural-language dialogue. The user describes the desired simulation in plain English, and a team of specialized agents collaborates to produce a schema-valid, physically sane GEOS XML file, backed by real fluid-property and meshing computations rather than by free-form text generation. Figure \ref{fig:agents_scheme} illustrates the overall functioning of Agents4GEOS.

The remainder of the paper is organized as follows. The next section details the specialized agents and their supporting tooling: the agent registry and its cost-aware model routing, the domain-aware MCP tool suite, the knowledge modules that encode GEOS expertise, and the coordination patterns through which the agents collaborate. We then demonstrate Agents4GEOS in action: first by following one real session end to end, in which the system reproduces a published PUNQ-S3 CO$_2$ injection benchmark from a one-sentence request, and then by validating the GEOS datasets generated by the system against published ECLIPSE and MRST results and presenting forecasts from the Plumecast GNN surrogate trained on those datasets. We close with conclusions, current limitations, and directions for future work.

%%%%%%%%%%%%%%%%%%%%%%%%%%%%%%%%%
\section{Agents4GEOS Description}\label{sec:description}

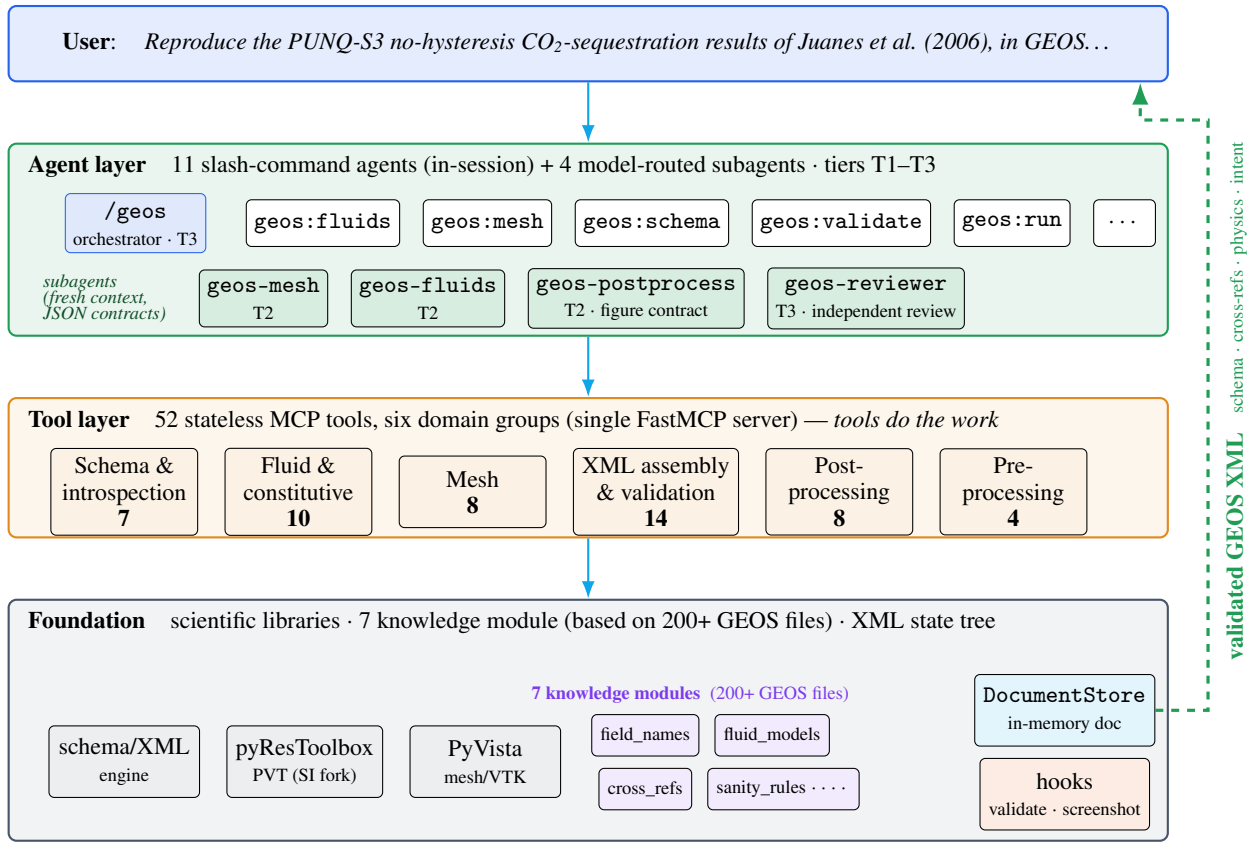
\begin{figure*}[htbp]
\centering
\resizebox{\linewidth}{!}{%
\begin{tikzpicture}[
  font=\small, >=Latex, node distance=4mm,
  band/.style={draw, thick, rounded corners=3pt, align=center, minimum height=1.0cm},
  chip/.style={draw, rounded corners=2pt, fill=white, font=\footnotesize, inner sep=3pt, align=center, minimum height=0.62cm},
  tool/.style={draw, rounded corners=2pt, fill=toolorange!12, font=\footnotesize, align=center, minimum height=0.95cm, minimum width=1.95cm},
  found/.style={draw, rounded corners=2pt, fill=foundgray!10, font=\footnotesize, align=center, minimum height=0.95cm},
  know/.style={draw, rounded corners=2pt, fill=knowpurple!10, font=\scriptsize, align=center, minimum height=0.55cm},
  lbl/.style={font=\footnotesize\bfseries, align=left},
  down/.style={->, thick, flowcyan},
]
% ---------- L1 USER ----------
\node[band, fill=userblue!12, draw=userblue, minimum width=15.4cm] (user)
  {\textbf{User}: \;\;
   {\footnotesize\itshape Reproduce the PUNQ-S3 no-hysteresis CO\textsubscript{2}-sequestration results of Juanes et al. (2006), in GEOS\ldots}};
% ---------- L2 AGENTS ----------
\node[band, fill=agentgreen!10, draw=agentgreen, minimum width=15.4cm, below=8mm of user, minimum height=2.55cm] (agentband) {};
\node[lbl, anchor=north west] at ($(agentband.north west)+(0.15,-0.05)$) {Agent layer \;\; {\footnotesize\mdseries 11 slash-command agents (in-session) + 4 model-routed subagents $\cdot$ tiers T1--T3}};
\node[chip, fill=userblue!14, draw=userblue] (orch) at ($(agentband.west)+(1.7,0.22)$) {\texttt{/geos}\\{\scriptsize orchestrator $\cdot$ T3}};
\node[chip] (a1) at ($(orch.east)+(1.55,0)$) {\texttt{geos:fluids}};
\node[chip, right=3mm of a1] (a2) {\texttt{geos:mesh}};
\node[chip, right=3mm of a2] (a3) {\texttt{geos:schema}};
\node[chip, right=3mm of a3] (a4) {\texttt{geos:validate}};
\node[chip, right=3mm of a4] (a5) {\texttt{geos:run}};
\node[chip, right=3mm of a5] (a6) {\;$\cdots$\;};
\node[font=\scriptsize\itshape, align=left, anchor=west, text=agentgreen!50!black] (sublbl) at ($(agentband.west)+(0.35,-0.78)$) {subagents\\[-2pt](fresh context,\\[-2pt]JSON contracts)};
\node[chip, fill=agentgreen!18] (s1) at ($(agentband.west)+(3.4,-0.78)$) {\texttt{geos-mesh}\\{\scriptsize T2}};
\node[chip, fill=agentgreen!18, right=3mm of s1] (s2) {\texttt{geos-fluids}\\{\scriptsize T2}};
\node[chip, fill=agentgreen!18, right=3mm of s2] (s3) {\texttt{geos-postprocess}\\{\scriptsize T2 $\cdot$ figure contract}};
\node[chip, fill=agentgreen!18, right=3mm of s3] (s4) {\texttt{geos-reviewer}\\{\scriptsize T3 $\cdot$ independent review}};
% ---------- L3 TOOLS ----------
\node[band, fill=toolorange!8, draw=toolorange, minimum width=15.4cm, below=8mm of agentband, minimum height=1.85cm] (toolband) {};
\node[lbl, anchor=north west] at ($(toolband.north west)+(0.15,-0.05)$) {Tool layer \;\; {\footnotesize\mdseries 52 stateless MCP tools, six domain groups (single FastMCP server) --- \itshape tools do the work}};
\node[tool] (t1) at ($(toolband.west)+(1.55,-0.32)$) {Schema \&\\introspection\\\textbf{7}};
\node[tool, right=3.5mm of t1] (t2) {Fluid \&\\constitutive\\\textbf{10}};
\node[tool, right=3.5mm of t2] (t3) {Mesh\\\textbf{8}};
\node[tool, right=3.5mm of t3] (t4) {XML assembly\\\& validation\\\textbf{14}};
\node[tool, right=3.5mm of t4] (t5) {Post-\\processing\\\textbf{8}};
\node[tool, right=3.5mm of t5] (t6) {Pre-\\processing\\\textbf{4}};
% ---------- L4 FOUNDATION ----------
\node[band, fill=foundgray!7, draw=foundgray, minimum width=15.4cm, below=8mm of toolband, minimum height=3.2cm] (foundband) {};
\node[lbl, anchor=north west] at ($(foundband.north west)+(0.15,-0.05)$) {Foundation \;\; {\footnotesize\mdseries scientific libraries $\cdot$ 7 knowledge module (based on 200+ GEOS files) $\cdot$ XML state tree}};
\node[found, minimum width=2.0cm] (lib1) at ($(foundband.west)+(1.55,-0.55)$) {schema/XML\\{\scriptsize engine}};
\node[found, minimum width=2.0cm, right=3.5mm of lib1] (lib2) {pyResToolbox\\{\scriptsize PVT (SI fork)}};
\node[found, minimum width=2.0cm, right=3.5mm of lib2] (lib3) {PyVista\\{\scriptsize mesh/VTK}};
\node[know, right=4mm of lib3, yshift=3.5mm] (k1) {field\_names};
\node[know, right=2mm of k1] (k2) {fluid\_models};
\node[know, below=1.5mm of k1] (k3) {cross\_refs};
\node[know, right=2mm of k3] (k4) {sanity\_rules $\cdot\cdots$};
\node[knowpurple, font=\scriptsize\bfseries, above=0.3mm of k1, xshift=6mm] {7 knowledge modules \;{\scriptsize\mdseries(200+ GEOS files)}};
\node[found, fill=Red!8, minimum width=2.1cm, anchor=south east] (hooks) at ($(foundband.south east)+(-0.25,0.15)$) {hooks\\{\scriptsize validate $\cdot$ screenshot}};
\node[found, fill=flowcyan!12, minimum width=2.1cm, above=1.5mm of hooks] (store) {\texttt{DocumentStore}\\{\scriptsize in-memory doc}};
% ---------- DOWNWARD FLOW ----------
\draw[down] (user)   -- (agentband);
\draw[down] (agentband) -- (toolband);
\draw[down] (toolband)  -- (foundband);
% ---------- RETURN FLOW (validated XML) ----------
\coordinate (retx) at ($(store.east)+(0.7,0)$);
\coordinate (rety) at ($(user.east)!0.55!(foundband.east)$);
\coordinate (retentry) at ($(user.south east)+(-0.38,0)$);
\draw[->, very thick, agentgreen, dashed]
  (store.east) -- (retx) -- ($(retx |- retentry)+(0,-0.5)$) -| (retentry);
\node[rotate=90, anchor=north, font=\footnotesize\bfseries, text=agentgreen]
  at ($(retx |- rety)+(0.12,0.2)$)
  {validated GEOS XML \;\;{\scriptsize\mdseries schema $\cdot$ cross-refs $\cdot$ physics $\cdot$ intent}};
\end{tikzpicture}%
}
\caption{Layered architecture of Agents4GEOS. A natural-language request flows down through the agent and tool layers to the scientific foundation, and a schema-valid, physics-checked GEOS XML file flows back to the user (dashed). The agent layer combines eleven slash-command agents with four fresh-context subagents that return structured JSON validated against typed contracts; agents make decisions, the 52 stateless MCP tools (51 domain tools in six groups plus a server health check) do the work, knowledge modules encode domain expertise, and the in-memory \texttt{DocumentStore} holds the document (XML tree state) being assembled across successive tool calls.}
\label{fig:a4g-arch}
\end{figure*}

Agents4GEOS is implemented as a Model Context Protocol (MCP)\footnote{\url{https://modelcontextprotocol.io}} server exposed to Claude Code\footnote{\url{https://docs.anthropic.com/en/docs/claude-code}}, the agent \emph{harness} that operates the agentic loop, complemented by slash-command \emph{skills} that define agent roles, by fresh-context \emph{subagents} that the orchestrator dispatches for parallel computation and independent review, and by \emph{hooks} that react automatically to tool calls --- for instance, validating an XML file as soon as it is saved, or flagging newly written VTK output for visualization. Agentic workflows come with a vocabulary of their own, borrowed from software engineering rather than from the geosciences. Table~\ref{tab:a4g-glossary} collects the terms used in this paper; the remainder of this section details how each ingredient is realized in Agents4GEOS. Figure~\ref{fig:a4g-arch} summarizes the layered organization of the system.

The architecture is organized in four layers with a strict separation of concerns. At the top, the \textbf{user} interacts only through natural language. The \textbf{agent layer} comprises eleven slash-command agents --- a single orchestrator (\texttt{/geos}) that converses with the user, and ten domain specialists (schema, edit, validate, fluids, mesh, relperm, inspect, run, post-processing, and error curation) --- together with four \emph{fresh-context subagents} (mesh, fluids, post-processing, and an independent reviewer) that the orchestrator dispatches for parallel computation and quality control. Agents embody decisions --- what to do and in which order --- but perform no computation themselves. The \textbf{tool layer} provides 52 stateless MCP tools, grouped into six domains, which carry out all concrete work. Finally, the \textbf{foundation layer} couples an in-repository GEOS schema/XML engine (schema parsing, XML reading and writing, templates, and validation --- originally developed as the standalone geos-tui project and since adopted into the code base to make the system self-contained) with two established scientific libraries --- the SI fork of pyResToolbox (fluid PVT, relative permeability, and well performance) and PyVista (mesh creation and VTK input/output), complemented by the cmcrameri perceptually uniform colormaps used for publication-quality figures --- together with a set of knowledge modules distilled from an audit of more than $200$ official GEOS input files, and with an in-memory \texttt{DocumentStore} that holds the XML document being assembled across successive tool calls.

This separation follows a single guiding principle: \emph{tools do the work, knowledge encodes domain expertise, and agents decide what work to do}. Tools are stateless and unaware of one another; domain patterns - valid boundary-condition field names per solver, constitutive-model assemblies, attribute cross-references, and physics sanity heuristics - live exclusively in the knowledge modules, so that a newly discovered pattern is added in a single place rather than scattered across the code base.

\begin{table*}[ht!]
\centering
\caption{Glossary of agentic-workflow terms as used in this paper.}
\label{tab:a4g-glossary}
\vspace{5pt}
\begin{tabular}{l p{0.72\linewidth}}
\hline
\textbf{Term} & \textbf{Meaning} \\
\hline
Agent & A large-language-model instance configured for a narrow role: it plans, decides which tools to call and in which order, and interprets the results, but performs no numerical work itself. \\
Orchestrator & The single agent that converses with the user, decomposes the request, and dispatches the specialists (\texttt{/geos}). \\
Skill & A version-controlled text file that defines an agent: its role, the tools it may call, its knowledge dependencies, and its capability tier; invoked as a slash command such as \texttt{/geos:fluids}. \\
Subagent & An agent dispatched as an independent instance with a \emph{fresh context} --- it shares no conversation history with its dispatcher --- that returns structured JSON validated against a typed contract, never free-form prose. \\
Tool & A stateless, deterministic function exposed to agents through MCP; tools carry out all concrete computation (fluid properties, meshing, validation, rendering). \\
MCP & Model Context Protocol, an open standard through which a language-model runtime discovers and calls external tool servers. \\
Hook & A script triggered automatically by a runtime event (e.g., a file being written), outside the model's control; used here to validate XML on save and to flag new VTK output. \\
Harness & The runtime that operates the agentic loop --- assembling context, calling the model, executing the requested tool calls, and enforcing permissions; Claude Code in this work. \\
Knowledge module & A versioned data file encoding domain expertise (valid field names, constitutive assemblies, sanity rules) that tools consult, keeping such patterns out of both prompts and tool code. \\
Capability tier & The complexity class of an agent's task (retrieval, synthesis, planning). Fresh-context subagents are routed to the cheapest language model that reliably handles their tier; in-session agents inherit the orchestrator's model, and their tier documents complexity. \\
\hline
\end{tabular}
\end{table*}

\subsection{Agent Registry and Model Routing}

Each agent is defined by a skill file that fixes its role, the subset of tools it may call, its knowledge dependencies, and a \emph{capability tier} reflecting the cognitive complexity of its task: Tier~1 (retrieval: lookup, extraction, formatting), Tier~2 (synthesis: combining data, XML assembly, validation, property computation), and Tier~3 (planning: multi-step reasoning, user dialogue, debugging). Tiers express a cost-aware model-routing policy --- use the cheapest model that can reliably handle the task, and tier up whenever a cheap model would be unreliable, since a wrong answer costs more than the savings. The policy can only be \emph{enforced}, however, where a model choice exists. The eleven slash-command agents run as skills inside the orchestrator's session and therefore inherit its (Tier-3) model; for them, the tier documents the complexity of the task rather than selecting a model. This is precisely what motivated promoting the most delegable specialists to explicit \emph{fresh-context subagents}: a subagent is dispatched as an independent agent instance whose definition pins the model matching its tier, so mesh and fluid computations actually run on a mid-sized model while the independent review runs on the most capable one. As language models improve, a subagent may move down a tier without any change to its definition. Table~\ref{tab:a4g-agents} lists the agents, their tiers, and their roles.

The four fresh-context subagents start from a clean context, sharing no conversation history with the orchestrator: \texttt{geos-mesh} and \texttt{geos-fluids} compute mesh and fluid-model fragments, \texttt{geos-postprocess} analyzes simulation output, and \texttt{geos-reviewer} performs an independent review of every assembled deck --- schema validity, cross-references, physics realism, and, crucially, \emph{fidelity to the user's stated intent} (injection rates, run duration, domain dimensions, output cadence), which it can judge precisely because it shares none of the builder's context. Each subagent returns structured JSON that is validated against a typed contract (\texttt{MeshResult}, \texttt{FluidResult}, \texttt{PostprocessResult}, \texttt{ReviewFinding}) before the orchestrator applies it; a result that violates its contract is rejected programmatically, and the orchestrator falls back to computing the fragment inline.

\begin{table*}[ht!]
\centering
\caption{The eleven Agents4GEOS slash-command agents (top) and four fresh-context subagents (bottom), their capability tiers, and roles. Tier~1 (retrieval), Tier~2 (synthesis), and Tier~3 (planning) map to progressively more capable language models; the mapping is enforced for the subagents, whose definitions pin the model of their tier, whereas the slash-command agents run in the orchestrator's session and their tier records task complexity.}
\label{tab:a4g-agents}
\vspace{5pt}
\begin{tabular}{l c l}
\hline
\textbf{Agent} & \textbf{Tier} & \textbf{Role} \\
\hline
\texttt{/geos} (orchestrator) & 3 & Entry point; create, edit, and query simulations \\
\texttt{geos:run} & 3 & Run GEOS, analyze output, diagnose runtime errors \\
\texttt{geos:edit} & 2 & Load, modify, validate, and save XML \\
\texttt{geos:validate} & 2 & Schema + cross-reference + physics sanity checks \\
\texttt{geos:fluids} & 2 & Fluid PVT properties and model recommendation \\
\texttt{geos:mesh} & 2 & Create and visualize meshes \\
\texttt{geos:relperm} & 2 & Relative-permeability and capillary-pressure curves \\
\texttt{geos:postprocess} & 2 & Analyze VTK output (fields, evolution, balances) \\
\texttt{geos:schema} & 1 & Query the GEOS XSD schema \\
\texttt{geos:inspect} & 1 & Summarize the contents of an XML file \\
\texttt{geos:curate-errors} & 1 & Curate runtime-error logs for pattern learning \\
\hline
\texttt{geos-reviewer} & 3 & Independent deck review: schema, physics, intent fidelity \\
\texttt{geos-mesh} & 2 & Fresh-context mesh computation (\texttt{MeshResult}) \\
\texttt{geos-fluids} & 2 & Fresh-context fluid/PVT computation (\texttt{FluidResult}) \\
\texttt{geos-postprocess} & 2 & Post-run analysis under a publication-figure contract \\
\hline
\end{tabular}
\end{table*}

The orchestrator (\texttt{/geos}) is the only agent that converses with the user; the specialists report \emph{structured} results (dictionaries and tables, never free-form prose) back to it. This contract makes handoffs deterministic and keeps the dialogue coherent even when several specialists run concurrently.

\subsection{Domain-Aware Tool Suite}

The 52 MCP tools comprise 51 domain tools organized into six groups, plus a server health check, each group backed by a dedicated scientific library so that every quantity returned to the user results from an actual computation rather than from text generation. Table~\ref{tab:a4g-tools} summarizes the groups. The schema and XML groups rest on a schema/XML engine that originated as \emph{geos-tui}, a standalone text user interface for GEOS developed by our group, and was later adopted into the Agents4GEOS code base so that the system is self-contained.

\begin{table*}[ht!]
\centering
\caption{The six tool groups exposed by the Agents4GEOS MCP server, totaling 52 tools together with a server health check. Each group is backed by a dedicated scientific library or by the project's knowledge modules.}
\label{tab:a4g-tools}
\vspace{5pt}
\begin{tabular}{l c l}
\hline
\textbf{Tool group} & \textbf{\# tools} & \textbf{Backed by} \\
\hline
Schema \& introspection      & 7  & in-repo schema engine \\
Fluid \& constitutive        & 10 & pyResToolbox (SI units) \\
Mesh                         & 8  & PyVista \\
XML assembly \& validation   & 14 & in-repo XML engine + \texttt{xmllint} \\
Post-processing              & 8  & PyVista + pyResToolbox + cmcrameri \\
Preprocessing                & 4  & knowledge modules \\
Server utilities             & 1  & FastMCP \\
\hline
\textbf{Total}               & \textbf{52} & \\
\hline
\end{tabular}
\end{table*}

Representative tools illustrate the range of work delegated to this layer: \texttt{recommend\_fluid\_model} maps a natural-language scenario to a solver and a complete constitutive assembly; \texttt{compute\_brine\_properties} returns density, viscosity, and formation-volume factor in SI units; \texttt{generate\_internal\_mesh\_xml} emits a GEOS \texttt{InternalMesh} block from a domain specification; \texttt{validate\_cross\_references} checks that every internal name reference in the document resolves to an existing target; and \texttt{screenshot\_field} produces publication-quality renderings of VTK output fields.

\subsection{Knowledge Modules}

Domain expertise is encoded in seven Python knowledge modules --- \emph{field names} (solver $\rightarrow$ valid boundary/initial-condition fields), \emph{fluid models} (natural-language keywords $\rightarrow$ solver and constitutive assembly), \emph{cross-references} (attribute $\rightarrow$ target section), \emph{sanity rules} (physics and structural heuristics), \emph{unit conventions} (SI handling and bracketed unit expressions), \emph{formatting conventions} (canonical XML style), and \emph{preprocessing rules} (parameter expansion and include resolution). They are complemented by a two-stage \emph{example catalog} distilled from an audit of more than 200 official GEOS input files: a routing table that maps the user's intent to one of eight physics categories (single-phase, thermal, \CO2--brine, black oil, dead oil, compositional, immiscible, and wells), each with a detail file of decision rules and recommended starter decks. A curated prose record of runtime-error lessons completes the knowledge base. These modules are the single source of truth for domain patterns: tools read from them, and when a new pattern is discovered (for example, through automated curation of the runtime-error log by the \texttt{geos:curate-errors} agent) it is added to the appropriate module rather than hard-coded into a tool. This closes a lightweight learning loop in which lessons from past runs improve the assembly and validation of future simulations.

\subsection{Orchestration and Coordination}

\begin{figure*}[htbp]
\centering
\resizebox{\linewidth}{!}{%
\begin{tikzpicture}[scale=1.0, font=\footnotesize, >=Latex,  box/.style={draw, thick, rounded corners, align=center, inner sep=5pt}, io/.style={box, fill=yellow!18},   orch/.style={box, fill=blue!12}, agent/.style={box, fill=green!14, minimum width=2.6cm}, rev/.style={box, fill=red!10}, outbox/.style={box, fill=gray!16}]
\node[io] (req) {Natural-language\\ request};
\node[orch, right=1.0cm of req] (orch) {\texttt{/geos}\\ orchestrator\\ {\scriptsize Tier 3 $\cdot$ catalog routing}};
\node[agent, right=1.0cm of orch, yshift=0.85cm] (fluids) {\texttt{geos-fluids}\\ {\scriptsize Tier 2 $\cdot$ \texttt{FluidResult}}};
\node[agent, right=1.0cm of orch, yshift=-0.85cm] (mesh) {\texttt{geos-mesh}\\ {\scriptsize Tier 2 $\cdot$ \texttt{MeshResult}}};
\node[orch, right=5.2cm of orch] (asm) {assemble XML\\ + \texttt{geos:validate}\\ {\scriptsize in-session}};
\node[rev, right=1.0cm of asm] (rev) {\texttt{geos-reviewer}\\ {\scriptsize Tier 3 $\cdot$ fresh context}\\ {\scriptsize intent fidelity}};
\node[outbox, right=1.0cm of rev] (xml) {Validated \\ GEOS XML};
\draw[->, thick] (req) -- (orch);
\draw[->, thick] (orch.east) -- (fluids.west);
\draw[->, thick] (orch.east) -- (mesh.west);
\draw[->, thick] (fluids.east) -- (asm.west);
\draw[->, thick] (mesh.east) -- (asm.west);
\draw[->, thick] (asm) -- (rev);
\draw[->, thick] (rev) -- (xml);
\draw[->, thick, dashed] (rev.south)
  .. controls +(0,-1.4) and +(0,-1.4) ..
  node[below, font=\scriptsize, align=center]
    {structured findings $\rightarrow$ fix and re-review (max.\ 3 iterations)}
  (orch.south);
\end{tikzpicture}%
} % end of resizebox
\caption{End-to-end orchestration in Agents4GEOS for a simulation-creation request. The orchestrator routes the request through the example catalog to a physics category and dispatches two fresh-context compute subagents in parallel (\emph{fan-out}): \texttt{geos-fluids} selects a constitutive model and computes PVT properties, while \texttt{geos-mesh} builds the grid and its \texttt{InternalMesh} block; each returns structured JSON validated against a typed contract. The merged fragments are assembled into a document (\emph{pipeline}), which \texttt{geos:validate} checks against the schema, cross-references, and physics heuristics. The assembled deck is then handed to \texttt{geos-reviewer}, an independent fresh-context reviewer that also judges fidelity to the user's stated intent; on blocking findings, the orchestrator fixes the deck and dispatches a fresh reviewer (\emph{feedback loop}), up to three iterations, before returning the validated XML.}
\label{fig:a4g-workflow}
\end{figure*}
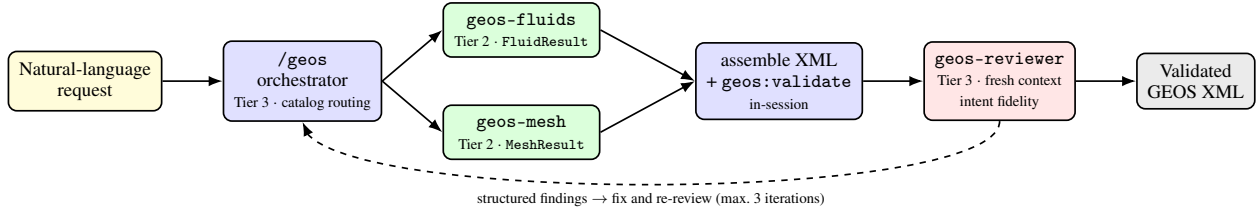

Agents4GEOS composes its agents through four recurring coordination patterns. In a \emph{pipeline}, one agent produces structured output that the next validates or transforms, halting rather than passing invalid data downstream. In a \emph{fan-out}, the orchestrator dispatches several independent fresh-context subagents in parallel and merges their structured results. In a \emph{feedback loop}, an agent's output is reviewed by an independent agent and, if deficient, the producer is re-invoked with corrective feedback, bounded by a maximum number of iterations (three in the current implementation). Finally, in a \emph{quality contract}, a subagent re-instantiated fresh on every dispatch enforces a non-negotiable output standard whose violation is detected programmatically --- \texttt{geos-postprocess}, for instance, must deliver publication-quality figures with perceptually uniform scientific colormaps (Crameri's \texttt{batlow} family), and any result using a banned colormap such as rainbow or jet is rejected before it reaches the user. Figure~\ref{fig:a4g-workflow} traces these patterns through a representative simulation-creation request.

Concretely, when a user asks for a \CO2 injection case, the orchestrator routes the request through the example catalog to the \CO2--brine physics category and fans out to the two compute subagents: \texttt{geos-fluids} recommends an appropriate flow model, for example \texttt{CO2BrinePhillipsFluid}, and computes the supporting PVT tables; \texttt{geos-mesh} generates a structured grid (in the current version of the code) together with its \texttt{InternalMesh} block and the geometry boxes for boundary conditions. The orchestrator merges these structured fragments, resolves the field names admitted by the chosen solver from the knowledge modules, assembles the document held in the \texttt{DocumentStore}, and hands it to \texttt{geos:validate}, which combines \texttt{xmllint} schema validation, cross-reference resolution, and physics sanity checks. The deck then faces \texttt{geos-reviewer}, which re-validates it with fresh eyes and checks every quantitative ask in the original request --- rates, durations, locations, output cadence --- before the validated file, ready to run in GEOS, is returned to the user. After a successful run, \texttt{geos-postprocess} is dispatched to summarize the output and render publication-quality figures under its colormap contract.

%%%%%%%%%%%%%%%%%%%%%%%%%%%%%%%%
\section{Agents4GEOS in Action}
\label{sect:agentsinaction}
Among many workflows that the described agents can execute, in this section one related to reproducibility is described in detail.

\subsection{Reproducing a published benchmark from a one-sentence request}

\begin{figure*}[ht!]
    \centering
    \includegraphics[width=0.9\linewidth]{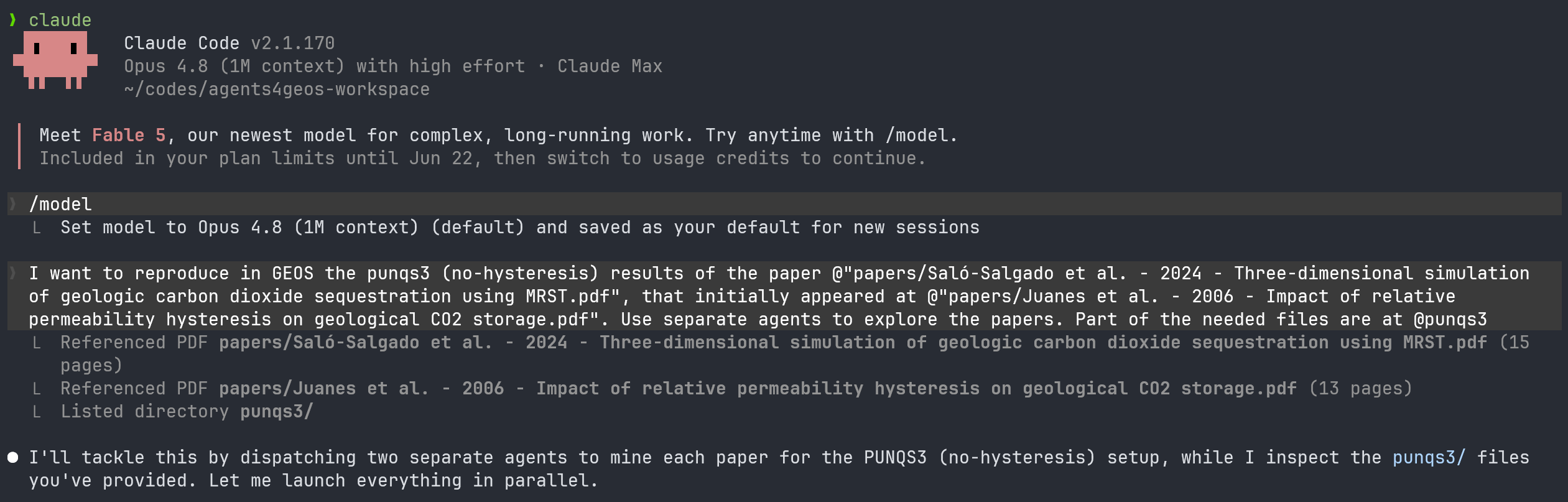}
    \caption{Plain English prompt for Agents4GEOS: A reservoir engineer asks Agents4GEOS to reproduce in GEOS the PUNQ-S3 no-hysteresis \CO2-sequestration results of \cite{SaloSalgado20242}.}
    \label{fig:initial_prompt}
\end{figure*}

\begin{figure}[ht!]
    \centering
    \includegraphics[width=\linewidth]{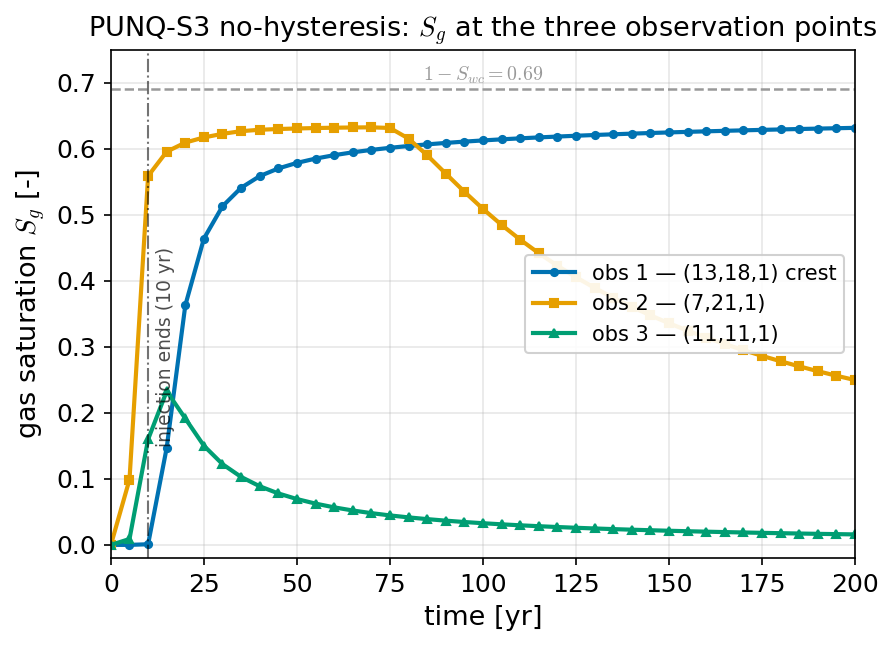}
    \caption{The payoff of the reproduction session: gas saturation versus time at the three fixed observation points of \cite{Juanes2006}, extracted by Agents4GEOS from the 101 snapshots of its agent-authored GEOS run. The crest cell (obs~1) accumulates toward the drainage endpoint $1-S_{wc}=0.69$ and holds, while the flank cells (obs~2, obs~3) spike during the 10-year injection and drain to near zero --- the no-hysteresis signature reported in both source papers (cf.\ Figure~6 of \cite{Juanes2006}).}
    \label{fig:punqs3-3obs}
\end{figure}

The clearest way to convey what Agents4GEOS does is to follow a real session end-to-end. A reservoir engineer asks, in plain English, to reproduce in GEOS the PUNQ-S3 no-hysteresis CO$_2$ sequestration results of \cite{SaloSalgado20242}, which in turn revisit the benchmark in \cite{Juanes2006}, providing only the two source PDF files and a set of mesh and well files. The initial prompt is shown in Figure \ref{fig:initial_prompt}  . Everything that follows, from mining the papers to the figure that closes the loop (Figure~\ref{fig:punqs3-3obs}), happened in a single conversational session, and took approximately one hour and consumed 343K tokens, in which the user wrote no XML and no shell commands. 

The orchestrator framed the problem as an extraction-then-translation task: the underlying physics is provided by the papers, and the work is mapping it onto GEOS's solver and constitutive vocabulary. It dispatched two paper-mining agents in parallel (a fan-out), each instructed to read one PDF file and return a quote-backed structured specification, while it inspected the provided files itself. The miners returned the grid numbers (a $19\times28\times5$ corner-point mesh with 1761 active cells), the fluid system (immiscible \CO2--brine at 40\,$^{\circ}$C), the Berea drainage relative permeabilities ($S_{wc}=0.31$, $S_{g,\max}=0.69$), the wells injection program (8 injectors operating with maximum bottom hole pressure of 160\,bar, along 10 years of injection followed by 490 years of migration), and, crucially, the fact that role of the relative permeability hysteresis model is explored by the two source articles. The local inspection goes beyond ingesting data; it identified a maximum porosity of 298 (non-physical) on the boundary cells, and recognized that it serves as the aquifer open boundary condition of \cite{Juanes2006} embedded in the mesh. It also recognized that the geological domain is located at a depth of $\approx$2340\,m, despite the benchmark relocating the reservoir to 840\,m to obtain 90\,bar of initial hydrostatic pressure, which allows the injected supercritical \CO2 to form plumes flowing towards the interior of the reservoir.

\begin{figure*}[ht!]
    \centering
    \includegraphics[width=0.9\linewidth]{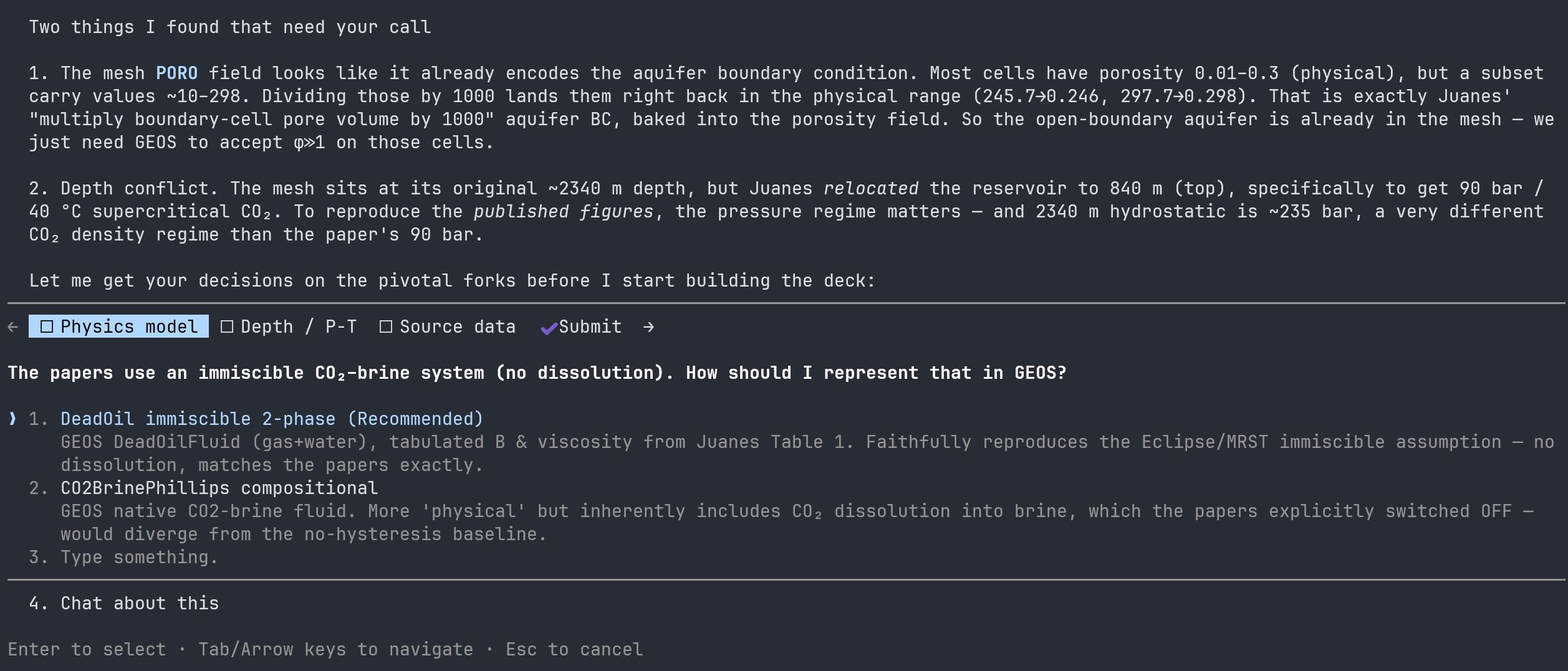}
    \caption{Agents4GEOS decision point:  \texttt{"The papers use an immiscible COz-brine system (no dissolution). How should I represent that in GEOS?".}}
    \label{fig:decision}
\end{figure*}

The system then returns to the user a three questions \textit{decision gate}, as shown in Figure \ref{fig:decision}, each carrying a recommendation and its trade-off. The recommendations being: 1. model the system as immiscible dead-oil (GEOS library denomination) rather than compositional \CO2--brine, allowing for gas dissolution; 2. displace the geological formation to 840\,m; and 3. reconstruct state equations PVT and relative-permeability inputs from the adopted source values. The user accepted all three recommendations. The depth question illustrates the level of reasoning involved: because the hydrostatic initial pressure depends on the spatial position of the top of the formation at 90\,bar. This relocation is assumed not to interfere with the exploration of the main aspects of the flow dynamics, especially in capturing the primary CO2 storage mechanisms. This is the human-in-the-loop contract of Agents4GEOS: the system decides everything it can from sources, code, and sensible defaults, and reserves the user's attention for genuine scientific choices.

The input deck build followed the \texttt{/geos} workflow of section \ref{sec:description}. The catalog knowledge base routed the request to the dead-oil model and selected an initial XML, a curated \CO2-storage GEOS example, that connects the reservoir--well and hydrostatic equilibrium, which was then transformed exclusively through MCP tool calls --- never hand-edited text --- into the target configuration, namely, dead-oil \CO2/brine PVT tables reconstructed from the paper, a Brooks--Corey fit to the published drainage endpoints (a reversible curve, hence inherently hysteresis-free), the final mesh with its imported porosity and permeability, eight rate-controlled injectors, and a 500-year schedule. The artifacts of the input deck were validated and then audited by \texttt{geos-reviewer} in a fresh context, which returned one needed fixing (the equilibrium initial condition's component names had to match the fluid's) plus the advisories it had been asked to scrutinize. A particular discrepancy was identified between the two articles sources. The injection rate:  ``18\,rm$^3$/day per well'' of \cite{SaloSalgado20242} and the ``0.15 pore volumes in 10 years'' of \cite{Juanes2006} disagree by more than an order of magnitude. Re-querying both paper-mining agents for verbatim quotes traced the tension to the boundary-inflated pore volume, and the system kept the literal value of its stated reproduction target, but reported the discrepancy to the user instead of silently picking a number, a behavior enforced by the skill.

GEOS rejected the first input deck with instructions for the simulation twice, that shows that even though the syntax is right and validated, some errors are only caught at running time, and this is where the learning loop earns its keep. The first failure was geometric conflict: the provided well coordinates traced the deep corner pillar of the slanted corner-point cells, placing perforations about a meter below the layer-5 cell centers; the agent parsed the mesh, computed each well column's true cell-center depth from the cell corner nodes, and redefined the wells as short vertical stubs centered there. The second was an API subtlety: GEOS's \texttt{DeadOilFluid} derives component names from its phase names and silently ignores user-supplied ones, so the initial condition referencing \texttt{co2} could not resolve. Both diagnoses were applied and \emph{persisted to the runtime-error knowledge base}, so the next deck benefits.  A 500-year eight-well run completed in 2 \, minutes 49 \, seconds, producing 101 reservoir snapshots.

\begin{figure}[ht!]
    \centering
    \includegraphics[width=\linewidth]{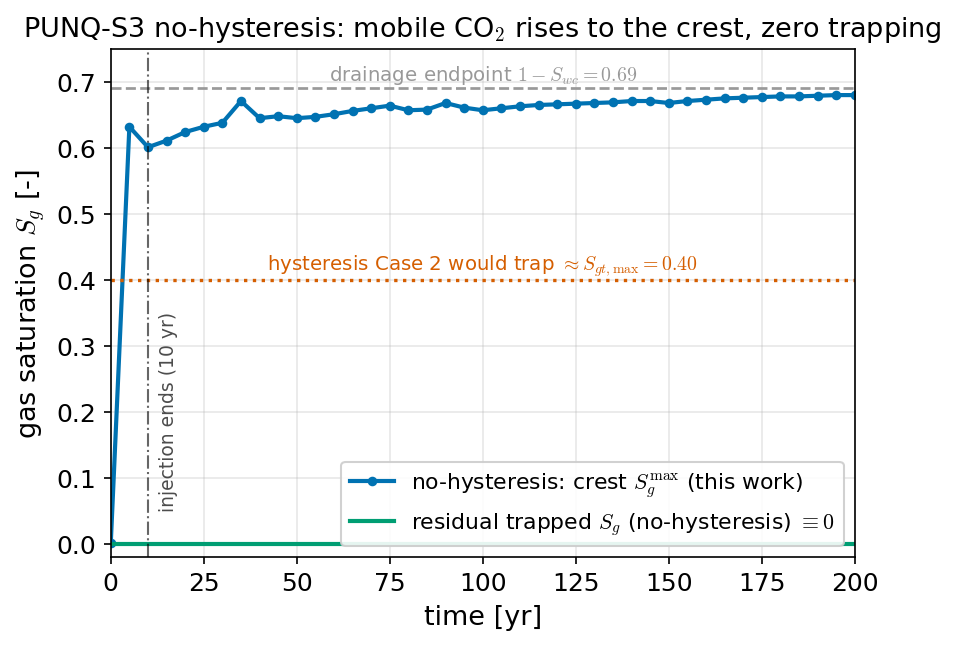}
    \caption{Initial gas saturation versus time graph produced by the subagent \texttt{geos-postprocess}, that took maximum over the domain as the observation variable.}
    \label{fig:punqs3-max-obs}
\end{figure}

Post-processing the results with  \texttt{geos-postprocess} subagent enables an overall evaluation of the  flow dynamics: gas migrates from the eight dispersed injectors, mixes with the resident brine, and forms a mobile plume at the top of the formation, with crest saturation approaching the drainage endpoint $1-S_{wc}=0.69$, residual trapping identically zero at all times, and a modest $\sim$6.4\,bar overpressure acting on the aquifer. The closing exchange is the part we find most instructive. The subagent produced Figure \ref{fig:punqs3-max-obs}, and the user asked which observation point a saturation-versus-time curve used; the orchestrator admitted it has been plotted the domain \emph{maximum}, a shortcut, then located the three fixed observation cells of \cite{Juanes2006} and extracted the true gas saturation at each across all 101 snapshots, as prompted by the user. Figure~\ref{fig:punqs3-3obs} is the result given by Agents4GEOS: the crest cell accumulates and holds, while the flank cells spike during injection and drain back to nearly zero, precisely the contrast the hysteresis benchmark exists to demonstrate, matching perfectly the results of \cite{Juanes2006}. Not everything is glossy, and we report it as the session did: the injection-rate tension remains unreconciled in the sources; the relative permeability is an endpoint fit rather than the raw published curve; and the plan-view saturation maps used a density-based proxy because the screenshot tool renders the norm of multi-component fields. In our view, a feature a scientific agent must have.

\subsection{Agents4GEOS supporting Graph Neural Networks (GNN) surrogates training}

\begin{figure*}
    \centering
    \includegraphics[width=0.9\linewidth]{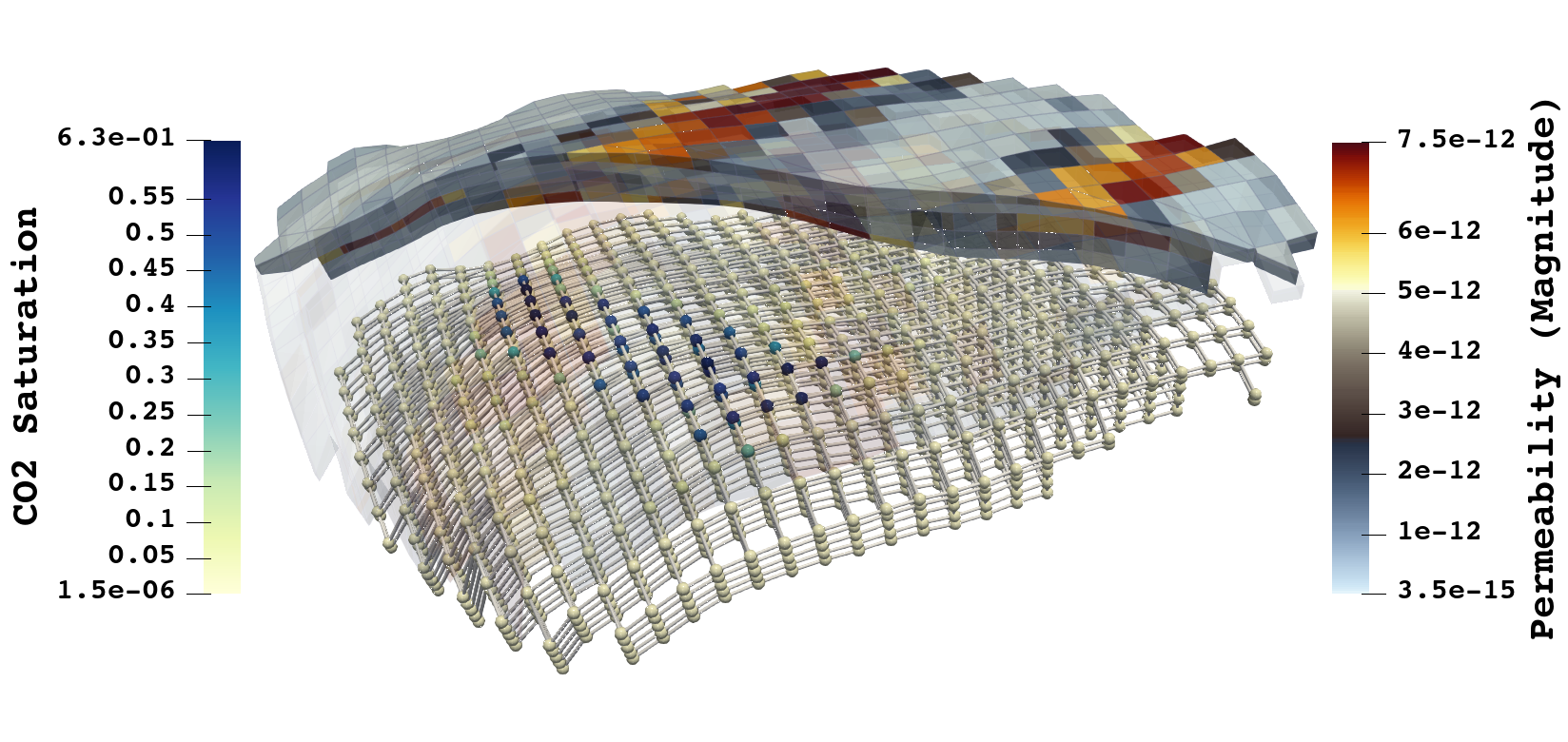}
    \caption{Top: the permeability spatial field of the PUNQ-S3 benchmark. Bottom: a representation of the graph structure extracted from the PUNQ-S3 mesh, where graph nodes corresponds to cells and edges their neighborhood relationship, and node colors the CO$_2$ saturation of the cell, one of the so called node features \cite{ju2024learning, luna2026benchmarking}.}
    \label{fig:CoverMesh}
\end{figure*}

A single validated simulation is in the core of a larger pipeline: running the same agent-built scaffold across an ensemble of parametric realizations, in our case permeability, yields physically consistent datasets on which GNN surrogates are trained. Agents4GEOS was used to generate 200 PUNQ-S3 simulations with sampled variations of the permeability spatial field of the PUNQ-S3 benchmark as represented at Figure \ref{fig:CoverMesh} (top). Also, it shows (bottom) the graph structure that serves as an input of \textit{Plumecast}, that is, another product of scientific agentic coding with human-in-the-loop.

\begin{figure}
    \centering
    \includegraphics[width=0.9\linewidth]{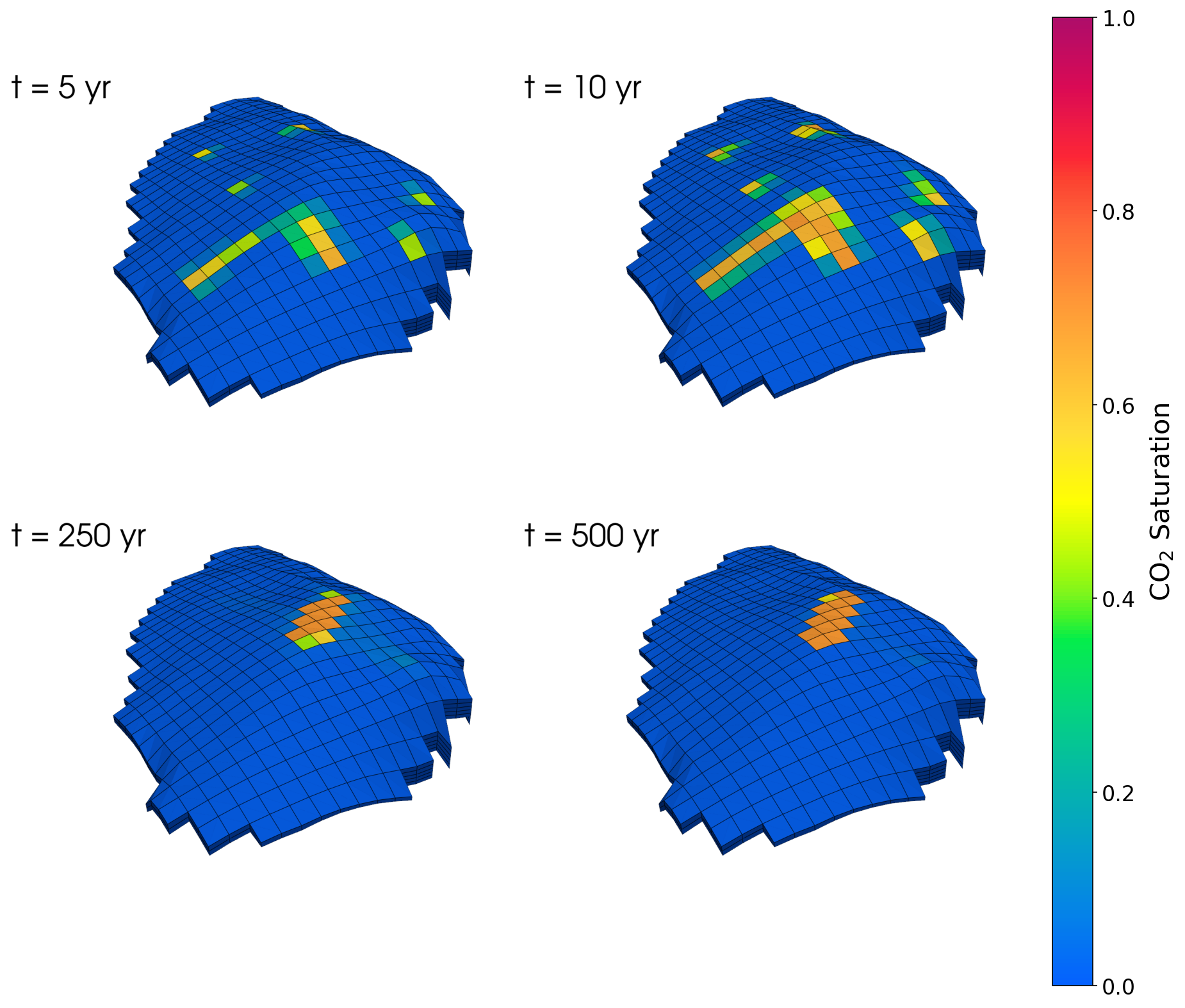}
    \caption{GEOS solutions for the PUNQ-3D non-hysteresis case from \cite{SaloSalgado20242} obtained with the dataset produced by Agents4GEOS.}
    \label{fig:nohysteresis}
\end{figure}

%%%%%%%%%%%%%%%%%%%%
\begin{figure}
    \centering
    \includegraphics[width=0.9\linewidth]{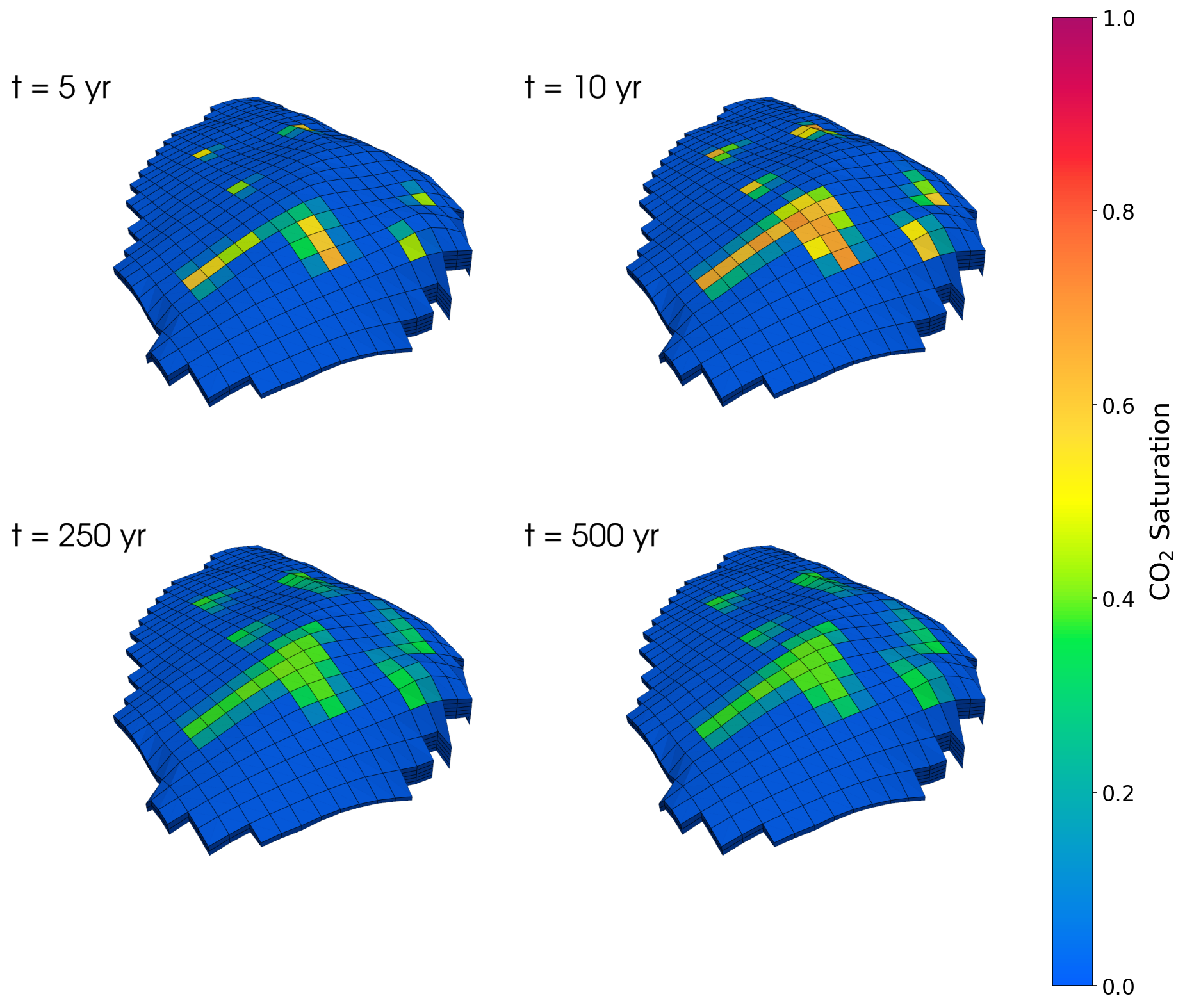}
    \caption{GEOS solutions for the PUNQ-3D hysteresis case from \cite{SaloSalgado20242} obtained with the dataset produced by Agents4GEOS.}
    \label{fig:hysteresis}
\end{figure}

%%%%%%%%%%%%%%%%%%%

Plumecast is a graph-based (GNN) surrogate built on top of PyG and PGT (ADD REFS). Its architecture is inspired by the model in \cite{ju2024learning} and the AnisoMeshGraph-LSTM model of \cite{luna2026benchmarking}. Plumecast combines a MeshGraphNet \cite{Pfaff2021MeshGraphNet} spatial processor with a GraphConv--LSTM temporal recurrence and is rolled out autoregressively over the time simulation horizon. Compared to AnisoMeshGraph-LSTM, the spatial processor uses isotropic message passing but is informed by reservoir physics through additional features: tabulated relative permeabilities from drainage curves, transmissibility-aware edge attributes derived from cell geometry and permeability, automatic detection of injection-well cells from the simulator output, and porosity consistent with Kozeny--Carman scaling. Training uses a constant curriculum, together with a loss that explicitly penalizes errors within the plume to suppress the false-spreading failure mode of regression-based surrogates.

\begin{figure*}[h]
    \centering
    \includegraphics[width=0.65\linewidth]{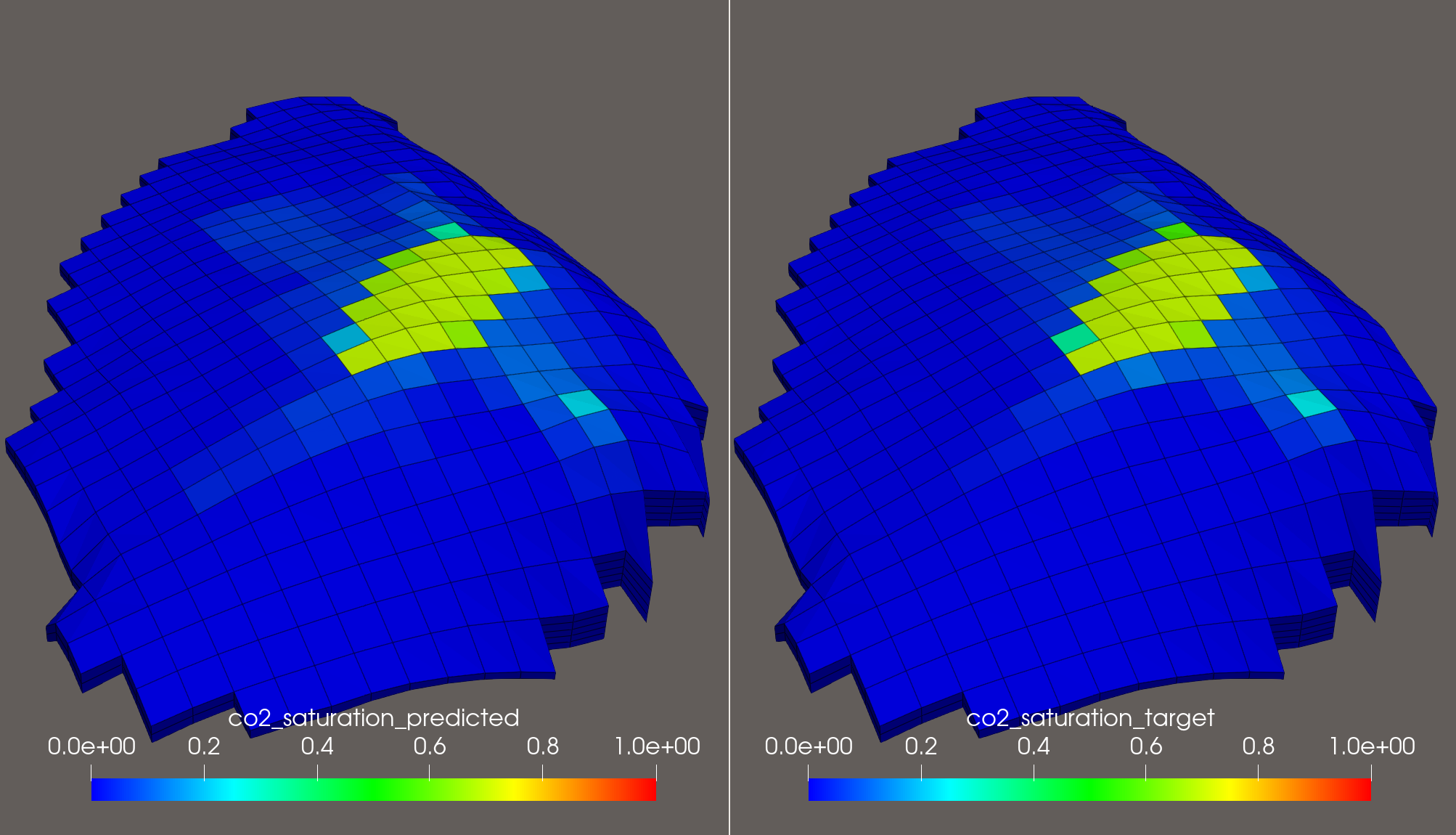}
    \caption{\CO2 saturation at $t=100$ years for the no hysteresis case obtained with  (left) Plumecast GNN surrogate and (right)  GEOS.}
    \label{fig:comparisonPlumecastAndJuanes}
\end{figure*}

In Figure~\ref{fig:comparisonPlumecastAndJuanes}, we present a comparison between the predicted \CO2 saturation field, for the case without hysteresis, at $t=100$ years, generated by Plumecast and the one computed by the high-fidelity GEOS simulator. Note the good visual agreement between the two \CO2 saturation spatial distributions.

%%%%%%%%%%%%%%

Finally, Fig. \ref{fig:CoverMesh} shows a view that links the spatial domain to the associated graph structure. The first is colored with the permeability values, while the latter is colored with CO2 saturation in each node.

%%%%%%%%%%%%%%%%%%%%%%%%%%%%%%%
\section{Discussion}

Agents4GEOS demonstrates that a multi-agent system organized around a strict separation of concerns — where agents plan, tools compute, and knowledge modules encode domain expertise — can substantially lower the barrier to managing GEOS input files. By grounding every quantity in actual scientific computations rather than free-form text generation, and by closing a lightweight learning loop through automated error curation, the system goes beyond prompt engineering to deliver physically consistent simulations from natural-language descriptions.

The PUNQ-S3 CO$_2$ injection dataset generated through this pipeline provides a controlled, high-quality ensemble of 100 training and 100 test simulations spanning a representative permeability range, with physically consistent porosity derived through the Kozeny–Carman relationship. Validation against published ECLIPSE and MRST results confirms the fidelity of the GEOS simulations produced by Agents4GEOS, lending confidence to datasets generated at scale through this automated workflow. The Plumecast GNN surrogate trained on this dataset further shows that the train-inference distribution mismatch inherent to autoregressive rollout can be effectively addressed through a constant-exposure curriculum combined with a composite loss that independently penalizes errors over the full domain and within the CO$_2$ plume. The resulting model achieves good agreement with GEOS ground-truth saturation fields at long horizons, confirming that graph-based surrogates informed by reservoir physics — transmissibility-aware edge attributes, tabulated relative permeabilities, and Kozeny–Carman-consistent porosity — can reproduce plume dynamics at a fraction of the computational cost of full simulation. Together, these contributions form an end-to-end pipeline from natural-language problem description to trained surrogate, establishing the infrastructure on which uncertainty quantification and many-query applications rely.

Despite this promise, several limitations remain. Meshing complex reservoirs is particularly challenging and warrants a dedicated specialized agent. While Agents4GEOS can already launch jobs on computing clusters, more robust skill coverage is needed to ensure all simulations complete correctly. Interacting with running simulations — monitoring convergence, diagnosing stalls, and steering or restarting runs mid-execution — remains an open direction. Beyond computational concerns, token usage has emerged as a meaningful factor in the overall budget, one that is no longer secondary to CPU or GPU hours.

The present framework is also the first stage of a broader roadmap. We are extending the agents toward a fully automated surrogate-construction workflow encompassing dataset generation, graph-neural-network architecture and hyperparameter selection, training management on GPU clusters, and uncertainty quantification. The aim is to reduce substantially the expertise required to build reliable surrogates for the many-query applications — history matching, optimization, and uncertainty quantification — that motivate the development of Agents4GEOS.

Finally, the broader question of how to evaluate agentic AI in this domain deserves attention. The Agents' Last Exam (ALE) benchmark \cite{sun2026ale} assesses agentic AI across 55 subfields and 13 industry clusters, yet only 9 of its 1,490 task instances are drawn from Mining, Petroleum, and Geological Engineering. This underrepresentation reflects the relative scarcity of verifiable, expert-sourced workflows in subsurface engineering compared to more digitally mature fields, and highlights a pressing need for richer, domain-specific benchmarks capable of properly assessing and driving progress in AI-assisted field development tasks.

%%%%%%%%%%%%%%%%%%%%%%%%%%%%%%%
\section{Conclusions}
Agentic workflows like Agents4GEOS are transforming the accessibility of high-fidelity open-source simulation tools such as GEOS. By reducing the learning curve for newcomers while enabling experts to build upon and extend existing capabilities, these workflows democratize access to sophisticated physical simulation engines. The use of contextual information provides superior token efficiency compared to open LLM approaches and facilitates the accumulation of institutional knowledge through user contributions, creating a sustainable and scalable framework for scientific software adoption.

%%%%%%%%%%%%%%%%%%%%%%%%%%%%%%%%
\section*{Acknowledgements}
This study was partially financed by the Coordenação de Aperfeiçoamento de Pessoal de Nível Superior-Brasil (CAPES)—Finance Code 001 and CNPq Brazil. This research was also sponsored by TotalEnergies E\&P Brazil under the 1\% ANP obligation (GNN CO2 Project 24630-6), and TotalEnergies EP R\&T US.

%%%%%%%%%%%%%%%%%%%%%%%%%%%%%%%
\section*{Conflict of Interest Statement}
The authors have no conflicts of interest to declare that are relevant to the content of this work.

% --- Bibliography ---
%\bibliographystyle{plain}
\bibliographystyle{unsrt}
\bibliography{references2}

\end{document}